# Application of Wiener-Hopf technique to linear non homogeneous integral equations for a new representation of Chandrasekhar's H- function in Radiative Transfer , its existence and uniqueness .

**Rabindra Nath Das***

Visiting Faculty  Member  , Heritage Institute of Technology , Chowbaga Road, Anandapur , P.O.East Kolkata Township, Kolkata-7000107, West Bengal , India
Emailaddress : rndas1946@yahoo.com

**Abstract:**  In this paper, the linear non homogeneous integral equation of H – functions is considered to find a new form of H-function as its solution .The Wiener –Hopf technique is used to express a known function into two functions with different zones of analyticity. The linear non homogeneous integral equation is thereafter expressed into two different sets of functions having the different zones of regularity.The modified form of Liouville's theorem is thereafter used ,   Cauchy's Integral formulae are used to determine functional representation over the cut region in a complex plane  .The new form of H- function is derived both for conservative and non-conservative cases .The existence of solution of linear non homogeneous integral equations  and its uniqueness are shown .  For numerical calculation of this new  H-function ,  a set of  useful formulae are derived both for conservative and non-conservative cases.

**Key words :** Radiative transfer ,  linear  non homogeneous Integral equations, Wiener-Hopf technique, Cauchy Integral formulae .

*Permanenet address : KB- 9 , Flat- 7 , Sector –3 , Saltb Lake city ( Bidhannagar) , Kolkata – 700098

## Introduction :

The equation of transfer which governs the radiation field in a medium is an integro-differential equation for the intensity of radiation . Most of the problems in radiation transport are essentially that of solving that equation of transfer. The emergent intensity from the bounding face and intensity of radiation at any optical depth of the plane-parallel homogeneous scattering semi-infinite atmosphere are found in terms of H –functions of Chandrasekhar [1]. In cases of diffuse reflection and transmission from the bounding faces in a homogeneous plane parallel atmosphere of finite thickness , Chandrasekhar [1] obtained solutions in terms of coupled X-Y functions . Chandrashekhar [1], Mullikin [2] and Das [3] showed that those X- Y functions can be expressed in terms of H- functions . These functions are also used to model the reflection of light from surface materials of celestial bodies such as planets , moon and asteroids . Those functions are also employed in studies of radiation transport in spherical clouds. A multitude of authors has contributed to creating an impressive body of knowledge on properties of H- functions ,.Hence H - functions are of considerable importance in the theory of radiative transfer in planetary and stellar atmosphere. Correct numerical evaluation to the desired accuracy of these functions depend however upon the availability of a suitable form . Chandrasekhar[1] tabulated the values of those functions for isotropic and anisotropic scattering using the non linear non homogeneous integral equations of those functions . There is still a need for further detailed numerical work on H- functions on a suitable form for being further accurately computed to any desired accuracy. Accordingly we have re examined the question of framing a new form the H -functions. The linear non homogeneous integral equations are considered. The wiener-Hopf technique and Cauchy integral formulae are used with the theory of analytic continuation to find a new form as a solution . The existence of this new form and its uniqueness are shown . From this point of view ,we believe that the form of solution presented here is sufficiently different from others in the literature of H – functions to warrant its communication .

Chandrasekhar [1] , Kourganoff [4] , Busbridge [5] used the non-linear non homogeneous integral equation ( NNIE) and the linear non homogeneous integral equations (LNIE) for such H- functions to obtain an integral representation of H-functions in terms of a known function . Fox [6] derived a linear non-homogeneous singular integral equation (LNSIE) of H- functions from NNIE of H-functions and obtained solution by the theory of singular integral equation of .Muskhelishvilli [7] treating it to be a Riemann Hilbert problem and found a form of H – functions not in a closed form.

Busbridge [8], Dasgupta [9], Dasgupta[10], Dasgupta [11] derived a family of solutions of that LNIE of H- functions and found two physically meaningful solution from such family of solutions to satisfy the NNIE of H – equations. But Busbridge [8] concluded that it was most unsafe to solve the LNIE for the H-functions in place NNIE of H – functions as there were a family of solutions .all satisfying the conditions at the origin and at infinity.

Zweifel [12], Siewert [13], Garcia and Siewert [14], Barichello and Siewert [15], and Barichello and Siewert [16] used the Eigen function method and they derived a new X- function and represented the H – function in terms of that X-function. Bergeat and Rutily [17], Bergeat and Rutily [18], Rutily and Bergeat [19] obtained a new form of H-functions by using Cauchy Integral equations and some auxiliary functions.

Mullikin [2] considered Chandrasekhar's X- Y functions. He used the theory of linear singular equations of Muskhelvilli [7] to obtain Fredholm equations for X- Y functions. For semi-infinite atmosphere, those Fredholm equations are solved as a limiting case to give this new representation of the H- function without any mathematical details.

Das [20] solved the equation for radiative transfer to a problem in semi-infinite atmosphere by the method of Laplace transform and Wiener- Hopf technique to obtain that form of H- function of Dasgupta [11]. Das[21] obtained the same new form using the theory of linear singular integral equations in a non homogeneous Riemann Hilbert Problem.

In this paper we show the existence of the solution LNIE of H - functions and obtain the new representation of H –function and its uniqueness using the Wiener-Hopf technique, Cauchys Integral formulae and theory of analytic continuation both for non conservative and conservative cases. Some useful formulae for calculation of moments of H – functions both for non conservative and conservative cases are also derived.

## Analysis :

The H-functions of Chandrasekhar [1] satisfy NNIE of the form

$$H(x) = 1 + x H(x) \int_0^1 U(t) H(t) \, dt / (t+x), \qquad 0 \le x \le 1 \qquad (1)$$

. Here $U(u)$, the Characteristic function which occurs in astrophysical contexts is assumed to be an even, real, nonnegative and bounded function in the interval $0 \le u \le 1$ and satisfies the condition

$$U_0 = \int_0^1 U(u) \, du \le 1/2. \qquad (2)$$

Fox [6] and Dasgupta [10] assumed that the functions $U(u)$ also satisfy. the Holder condition in the interval (0,1)

The case $U_0 = \frac{1}{2}$ is called the conservative case when there is no true absorption of radiation and efficiency of scattering is unity and the case $U_0 < \frac{1}{2}$ is called the non-conservative case when on each scattering, more radiation was not emitted than was incident.

Let $z = x + iy$, a complex variable and in the complex z plane cut along $(-1,1)$, $H(z)$ will satisfy the relation

$$(H(z) H(-z))^{-1} = T(z) \qquad (3)$$

Here, $T(z)$, a known dispersion function,

$$T(z) = 1 - 2 z^2 \int_0^1 U(u) \, du / (z^2 - u^2), \qquad (4)$$

has the following properties in conservative and non conservative cases:

i) $T(z)$ is analytic in the complex z plane cut along $(-1,1)$, ii) it has two branch points at $-1$ and $+1$, iii) $T(z) \to 1$ as $|z| \to 0$, iv) it has only two simple zeros at infinity when $U_0 = \frac{1}{2}$, v) only two real simple zeros at $z = +1/k$ and $z = -1/k$ where $k$ is real $0 < k \leq 1$ when $U_0 < \frac{1}{2}$, vi) $T(z) \to -C/z^2$ as $z \to \infty$ when $U_0 = \frac{1}{2}$, $C = 2 U_2$, a real positive constant where $U_r = \int_0^1 u^r U(u) \, du$, $r = 1,2,3 \ldots$ vii) $T(z) \to D = 1 - 2 U_0$ as $z \to \infty$ when $U_0 < 1/2$, $D$ is a real positive constant, viii) $T(z)$ is bounded on the entire imaginary axis, ix) $T(\infty) = D = 1 - 2 U_0 = (1 - \int_0^1 H(x) U(x) \, dx)^2 = 1/(H(\infty))^2$, when $U_0 < 1/2$, x) it is an even function of $z$, xi) $T(z)$ is negative as $z \to 1+0$ when $U_0 < \frac{1}{2}$.. (5).

On assumption of Holder condition on the function $U(u)$ in the interval $(0,1)$, $T(z)$ becomes singular and takes the form when z approaches to x in $(0,1)$

$$T_c(x) = 1 - 2 x^2 \int_0^1 (U(u) - U(x)) \, du / (x^2 - u^2)$$

$$- x U(x) \ln((1+x)/(1-x)), \qquad (6)$$

Fox [6] and Busbridge[8] considered the LNIE for $H(z)$ function in the complex z plane cut along $(-1,1)$ as

$$T(z) H(z) = 1 + z \int_0^1 U(u) H(u) \, du / (u - z) \qquad (7).$$

with constraints

    I)    when $U_0 < 1/2$

$$1 + \int_0^1 U(u) H(u) \, du / (uk - 1) = 0 \qquad (7a)$$

$$1 + \int_0^1 U(u) H(u) \, du / (uk + 1) = 0 \qquad (7b)$$

ii) when $U_0 = 1/2$

$$1 - \int_0^1 U(u) H(u) \, du = 0 \qquad (7c).$$

The properties ( i to x ) of T (z) in the complex z plane cut along (-1,1) stated in (5) may be used to solve the equation (7) for H(z) in the complex z plane using the Wiener-Hopf technique and theory of analytic function for conservative and non conservative cases .

We can therefore write

$$T(z) = Q(z)(1 - k^2 z^2) / (1 - z^2) \qquad (8)$$

where Q(z) is a new function having the following properties in the complex z plane cut along (-1,1)

I) Q(z) is an even function in z ,ii)It is analytic in the complex z plane cut along (-1,1), iii)does not have any zeros in the complex z plane cut along (-1,1),iv)Q(z) → $A/z^2$ when z → ∞ and $U_0 < ½$ ,where $A = D/k^2$ ,v)Q(z) → $-C/z^2$ when z → ∞ and $U_0 = ½$ ,vi)Q (iy) is positive along the entire imaginary axis for ( $-\infty < y < \infty$ ) so there is no variation in arg( Q(z) ) along the imaginary axis, vii)Q(z) has zeros at z =1 and at z = -1 ,viii)Q(z) → 1 as 1 z 1 → 0, ix) log Q(z ) may be defined that branch which is analytic in the z plane cut along (-1,1) such that log Q(z) = log C + O ($1/z^2$) when z→ ∞ and $U_0 = ½$ , [f (z)= O ($z^n$) if 1 f(z)/$z^n$ 1 < a constant ],x) log Q(z ) may be defined that branch which is analytic in the z plane cut along (-1,1) such that log Q(z) = log A + O ($1/z^2$) when z→ ∞ and $U_0 < ½$ ,xi)since points z = 1 and z = -1 are branch points of T(z) , the zeros of Q(z) do not introduce new branch points in the z plane cut along (-1,1) , xii) near z =0 ,Q(z) = 1 + O ($z^2$) , log Q(z) = O($z^2$) , Q(z) is bounded along the imaginary axis and log Q(z) /z is integrable along loop M enclosing z=0.

We now try to express log Q(z) as a contour integral in the z plane cut along (-1,1) using Cauchy's integral formulae in such a way that the integral along one part of the contour will represent log $Q_+(z)$ and the integral along the rest of the contour will represent log $Q_-(z)$ so that Q(z) can be represented as a quotient of two functions having different zone of representation in the complex z plane cut along (-1,1) with common zone of analyticity .

We shall find a relation between A,C. and D :We know that T(1/k)= 0 when $U_0 < ½$ , and hence

$T(1/k) = D - 2 U_2 O(k^2) - 2U_4 (k^4) - O(k^6)$ when $U_0 < ½$ and k is small ,

In the limit k→ 0 and $U_0$ → ½ , $A = D/k^2$ → $C = 2U_2$ .

for the case $U_0 < \frac{1}{2}$   $Q(z) = A - O(z^{-2})$ for large z ,

$$\log Q(z) = \log A + O(z^{-2}) \text{ for large } z ;$$

for the case $U_0 = \frac{1}{2}$    $Q(z) = C - O(z^{-2})$   for large z
$$\log Q(z) = \log C + O(z^{-2}) \text{ for large } z .$$

Following Kourganoff [4], let  w be a complex variable in the w plane cut along (-1,1). Let  M be a simple closed contour surrounding the cut along (–1 , 1) and let N be the circle |w| = R where R is so large that  N contains M . Let w = z  lie between M and N.

By Cauchy's integral formula , we get

$$\log Q(z) = (2\pi i)^{-1} \int_N \log Q(w) \, dw / (w - z)$$

$$- (2\pi i)^{-1} \int_M \log Q(w) \, dw / (w - z) \qquad (9)$$

where both the integrals  are taken in the counterclockwise sense.

Let on N , w is defined by  w = R exp (i θ)  ( 0 ≤ θ ≤ 2π )  and
therefore on  N ,  log Q(w) = log A + O ( $R^{-2}$ exp (- 2 i θ ) ) ,      (10)
where  R is large , for the case  $U_0 < 1/2$ .

$$(2\pi i)^{-1} \int_N \log Q(w) \, dw / (w - z) = \log A \qquad (11)$$

and   on N , log Q(w) = log C + O ( $R^{-2}$ exp (- 2 i θ ) )
where    R is large for the case  $U_0 = \frac{1}{2}$ ,

$$(2\pi i)^{-1} \int_N \log Q(w) \, dw / (w - z) = \log C \qquad (12)$$

Hence on letting R→∞  and using   equations (11) & (12) we have

$$\log Q(z) = \log A - (2\pi i)^{-1} \int_M \log Q(w) \, dw / (w - z) \text{ for the case } U_0 < \frac{1}{2} \quad (13)$$

$$\log Q(z) = \log C - (2\pi i)^{-1} \int_M \log Q(w) \, dw / (w - z) \text{ for the case } U_0 = \frac{1}{2} \quad (14)$$

Equations (13) and (14) can be written in a compact form   to cover both the cases  as

$$\log Q(z) = \log S - (2\pi i)^{-1} \int_M \log Q(w)\, dw / (w - z) \qquad (15)$$

where $S = A$ for the case $U_0 < \frac{1}{2}$, and $S = C$ for the case $U_0 = \frac{1}{2}$.

We now deform the simple closed contour M into two loops $M_+$ and $M_-$ keeping w = z outside the loops where $M_+$ starts from w=0 and goes below the cut along (0,1) and returns to w=0 above the cut along (0,1) in counterclockwise sense and $M_-$ starts from w=0 and goes above the cut along (-1,0) and returns to w=0 below the cut along (-1,0) also in counter clockwise sense to write the integral on the RHS of equation (15) in the form

$$(2\pi i)^{-1} \int_M \log Q(w)\, dw / (w - z)$$

$$= (2\pi i)^{-1} \int_{M+} \log Q(w)\, dw / (w - z) + (2\pi i)^{-1} \int_{M-} \log Q(w)\, dw / (w - z) \qquad (16)$$

Equation (15) with equation (16) becomes

$$\log Q(z) = \log S + \log Q_+(z) - \log Q_-(z) \qquad (17)$$

$$\log Q_+(z) = -(2\pi i)^{-1} \int_{M+} \log Q(w)\, dw / (w - z) \qquad (18)$$

$$\log Q_-(z) = (2\pi i)^{-1} \int_{M-} \log Q(w)\, dw / (w - z) \qquad (19)$$

provided the integrals (18) & (19) converge.

From (17)

we get $\quad Q(z) = S\; Q_+(z) / Q_-(z) \qquad (20).$

For the convergence of the integral in equations (18), (19), the only point on loops $M_+$ or $M_-$ at which difficulty for convergence of the integrals in equations (18) & (19) can arise is w=0 and it is easy to see that near w=0, $Q(w) = 1 + O(w^2)$, $\log Q(w) = O(w^2)$, $Q(w)$ is bounded along the imaginary axis
and $\log Q(w)/w$ is integrable along loop $M_-$. Following Kourganoff[4], we have to show the integrability of $\log Q(z)/z$ along $M_-$.

$$\text{Let } V = (2\pi i)^{-1} \int_{M_-} \log Q(w)\, dw / w \qquad (21)$$

We deform the loop $M_-$ of this integral (21) into i) the imaginary axis from 0 to i R, ii) along the semicircle $C_1$ defined by $w = R \exp(i\theta)$ ($\pi/2 \leq \theta \leq 3\pi/2$) and iii) the imaginary axis from $-iR$ to zero in the counterclockwise sense. The two parts of the integral along the imaginary axis cancel as $Q(w) = Q(-w)$, $Q(w)$ is bounded along the imaginary axis and there is no variation of $\arg(Q(z))$ along the imaginary axis, then the integral (21) gives

$$V = (2\pi)^{-1} \pi \log Q(\infty) \text{ when } R \to \infty$$

$$= (1/2) \log(S) = S^{1/2}. \qquad (22)$$

Hence the integrals in equations (18) and (19) do converge at w=0. We shall now go for the convergence of the integrals in equations (19) first. , Since the singularities and zeros of f(z) become the singularities of log f(z), it can be seen that $\log Q_+(z)$ and hence $Q_+(z)$ is an analytic function of z when z lies outside the loop $M_+$ which covers the cut along (0,1) of the w plane and $\log Q_-(z)$ and hence $Q_-(z)$ is also an analytic function of z when z lies outside the loop $M_-$ which covers the cut along (-1,0) of the w plane and they are having common zone of analyticity in the complex z plane. we shall go for convergence of $Q_-(z)$. In integral in equation(19), the point z lies outside the contour $M_-$. Following Kourganoff [4], let $w = r \exp(i\theta_1)$ and $z = r_1 \exp(i\theta_2)$ and we can choose $M_-$ in such a manner that it lies entirely in the sector defined as follows:
Let $|\theta_2| \leq \beta < \pi$,

$\beta + \delta \leq \theta_1 \leq 2\pi - \beta - \delta$, $0 < \delta < \pi - \beta$, $\delta \leq \theta_1 - \theta_2$,

where $|\log Q(w)/(w-z)| \leq \csc\delta |\log Q(w)/w|$ and therefore

$$|\log Q_-(z)| = |(2\pi i)^{-1} \int_{M_-} \log Q(w) \, dw/(w-z)|, \qquad (23)$$

$$\leq |(2\pi i)^{-1} \csc\delta \int_{M_-} \log Q(w) \, dw/w|$$

$$\leq \csc\delta \cdot V \qquad (24)$$

where V is given by equation (22).

Hence the integral (23) converges uniformly for $|\theta_2| \leq \beta < \pi$. It can be proved that

$$|\log Q_-(z)| \leq m_0 = V \csc\delta ; \qquad (25)$$

$$(k_0)^{-1} \leq |Q_-(z)| \leq k_0, \qquad (26)$$

where $k_0 = \exp(m_0)$. and therefore, the integral in equation (19) is convergent. We can similarly prove that $Q_+(z)$ is also bounded in a sector in which $M_+$ lies and the integral in equation (18) is also convergent

Hence the L NIE in equation (7) with expression of Q (z) in equation (8) can be written as

$$Q(z) H(z) = (1-z^2)/(1 - k^2 z^2) \left(1 + z \int_0^1 U(u) H(u) du / (u - z)\right). \quad (27)$$

Equation (27) and equation (20) after slight rearrangement can be written as

$$H(z)(1 + k z) S^{1/2} / ((1+z) Q_-(z)) = S^{-1/2} (1-z)\left(1 + z \int_0^1 U(u) H(u) du / (u-z)\right) / ((1-k z) Q_+(z)) \quad (28).$$

The left hand side of equation (28) is the product of functions which are analytic function of z when z lies outside the loop $M_-$ and right hand side of equation (28) is also the product of functions which are analytic function of z when z lies outside the loop $M_+$. Those two functions are equal at all points common to the domains and so each gives the analytic continuation of the other and together they represent a function G(z) which is regular in the whole z plane from which z =0 is excluded at present but infinity is included. As this G(z) would be a polynomial and it is bounded at the origin of the z plane, then by modified form of Liouville's theorem,

$$G(z) = T, \text{ a constant}. \quad (29)$$

We have to determine T. Equation (28) with equation (29) gives

$$H(z)(1 + k z) S^{1/2} / ((1+z) Q_-(z)) = T, \quad (30)$$

$$S^{-1/2} (1-z)\left(1 + z \int_0^1 U(u) H(u) du / (u - z)\right) / ((1-k z) Q_+(z)) = T, \quad (31)$$

From (30) we get

$$H(z) = T S^{-1/2} (1 + z) Q_-(z) / (1 + k z) \quad (32)$$

Following Busbridge [8], it can be seen from equation (1) that

 $| (H(x))^{-1} - 1 | \to 0$ when $|x| \to 0$

This shows that $H(x) \to 1$ as $x \to 0$.

Therefore we can take

$H(0) = 1$ \quad (33)

Equation (32) gives when | z | tends to zero. ,

$$H(0) = T S^{-1/2} Q_-(0) \quad (34)$$

where $Q_-(0) = (2\pi i)^{-1} \int_{M_-} \log Q(w) \, dw / w \quad (35)$

$$= V = S^{1/2} \quad [\text{using equation (22)}] \quad (36)$$

Hence from equations (30 - 36), we get

$$T = 1 \quad (37)$$

We calculate the exact expression of $\log Q_-(z)$: In equation (19), we may deform the contour $M_-$ in the following manner.: Let $M_-$ consist of i) a circle $C_0$: $w = r \exp(i\theta)$ about origin where $\theta$ runs from $-\pi$ to $\pi$, ii) the upper side of the cut along $(-1, 0)$ from $-r$ to $-1+r$, iii) the circle $C_{-1}$: $w = -1 + \exp(i\Phi)$ about $-1$ where $\Phi$ runs from $0$ to $2\pi$, iv) the lower side of the cut along $(-1,0)$ from $-1+r$ to $-r$ where $r$ is sufficiently small, all are in counterclockwise sense. $z$ is outside $M_-$.
 On $C_0$ the integral will be zero when $r$ is sufficiently small .as $\log Q(w) = O(w^2)$, when $|z| > 0$ $1/(w-z) = -z^{-1} + O(w)$ and $w$ is sufficiently small. On $C_{-1}$, $\log Q(w)$ is bounded and the integrand $\log Q(w)/(w-z)$ will tend to zero when $r$ is sufficiently small for all values of $z$ and the value of the integral will be zero. On upper side of the cut along $(-1,0)$ $w = -t$, $t$ is real, where $t$ runs from $0$ to $1$ and $0 < t < 1$, $\log Q(w)$ is denoted as $\log Q^U(-t)$ as $w$ is approaching the cut along $(-1,0)$ from above. On lower side of the cut along $(-1,0)$, $w = -t$, $t$ is real, where $t$ runs from $1$ to $0$ and $0 < t < 1$, $\log Q(w)$ is denoted as $\log Q^L(-t)$ as $w$ is approaching the cut along $(-1,0)$ from below. As $Q(z)$ is even, $\log Q^U(-t) = \log Q^U(t)$, $\log Q^L(-t) = \log Q^L(t)$. Therefore, the integrals along upper side of the cut along $(-1,0)$ and along lower side of the cut along $(-1,0)$) respectively are

$$(2\pi i)^{-1} \int_0^1 \log Q^U(t) \, dt / (t+z) \quad (38)$$

$$(2\pi i)^{-1} \int_1^0 \log Q^L(t) \, dt / (t+z) \quad (39)$$

Hence, taking all contribution over the deformed contour of $M_-$ together, we get for equation (19)

$$\log Q_-(z) = (2\pi i)^{-1} \int_0^1 \log (Q^U(t) - \log Q^L(t)) \, dt / (t+z) \quad (40)$$

It is easy to show that

$$\log (Q^U(t) - \log Q^L(t)) = \log (T^U(t) - \log T^L(t)) \quad (41)$$

Using the properties of T (z) on upper side and lower side of the cut along (-1,0) respectively

$$T^U(t) = T_c(t) - \pi i t U(t) \quad (42)$$
$$T^L(t) = T_c(t) + \pi i t U(t) \quad (43)$$

Equation (41) with equations (41-43) gives

$$\log(T^U(t) - \log T^L(t)) = -2i \tan^{-1}(\pi t U(t) / T_c(t)) \quad (44).$$

Hence equation (40) with equation (44) becomes

$$\log Q_-(z) = - \int_0^1 \theta(t) \, dt / (t+z) \quad (45)$$

$$Q_-(z) = \exp\left(- \int_0^1 \theta(t) \, dt / (t+z)\right) \quad (46)$$

where $\theta(t) = \pi^{-1} \tan^{-1}(\pi t U(t) / T_c(t)) \quad (47)$

Equation (45) therefore indicates that $\log Q_-(z)$ and hence $Q_-(z)$ is analytic in the z plane cut along (-1,0) including infinity and has no zero in the z plane cut along (-1,0). We shall now calculate the expression of $\log Q_+(z)$. We know from equation (18)

$$\log Q_+(z) = (2\pi i)^{-1} \int_M \log Q(w) \, dw / (w - z) \quad (48)$$

We may deform the contour $M_+$ in the following manner.
Let $M_+$ consist of I) a circle $C_0$: w= r exp(iθ) about origin where θ runs from 0 to $2\pi$, ii) the lower side of the cut along (0, 1) from r to 1 +r, iii) the circle $C_1$: w = 1 + exp(iψ) about 1 where ψ runs from -π to π, iv) the upper side of the cut along (0,1) from 1-r to r where r is sufficiently small, all are in counterclockwise sense. , z is outside $M_+$.
. On $C_0$ the integral will be zero when r is sufficiently small. $\log Q(w) = O(w^2)$, when |z| > 0 $1/(w-z) = -z^{-1} + O(w)$ and w is sufficiently small. On $C_1$, log Q(w) is bounded and the integrand log Q(w) / (w-z) will tend to zero when r is sufficiently small for all values of z and the value of the integral will be zero. On lower side of the cut along (0,1) w = t, t is real, where t runs from 0 to 1 and 0<t<1 , log Q(w) is denoted as $\log Q^L(t)$ as w is approaching the cut along (0,1) from below. On the upper side of the cut along (0,1) , w = t, t is real, where t runs from 1 to 0 and 0<t<1 , log Q(w) is denoted as $\log Q^U(t)$ as w is approaching the cut along (0,1) from above. The integrals along lower side of the cut along (0,1) and along upper side of the cut along (0,1)) respectively then become

$$(2\pi i)^{-1} \int_0^1 \log Q^L(t) \, dt / (t - z) \quad (49)$$

$$(2\pi i)^{-1} \int_1^0 \log Q^U(t) \, dt / (t-z) \qquad (50)$$

Hence taking all contribution over the deformed contour of $M_+$ together we get from equation (48)

$$\log Q_+(z) = -(2\pi i)^{-1} \int_0^1 \log Q^L(t) \, dt / (t-z) - (2\pi i)^{-1} \int_1^0 \log Q^U(t) \, dt / (t-z)$$

$$= -(2\pi i)^{-1} \int_0^1 \log (Q^L(t) - \log Q^U(t)) \, dt / (t-z) \qquad (51)$$

We can prove that

$$\log (Q^L(t) - \log Q^U(t)) = \log (T^L(t) - \log T^U(t)) \qquad (52)$$

Using the properties of $T(z)$ on upper side and lower side of the cut along $(0,1)$ respectively

$$T^U(t) = T_c(t) + \pi i t U(t) \qquad (53)$$
$$T^L(t) = T_c(t) - \pi i t U(t) \qquad (54)$$

Equation (52) with equation (53) and (54) gives

$$\log (T^L(t) - \log T^U(t)) = \log (T_c(t) - \pi i t U(t)) - \log (T_c(t) + \pi i t U(t)) \qquad (55)$$

Taking the principal branch of $\log (T_c(t) - \pi i t U(t))$ and $\log (T_c(t) + \pi i t U(t))$, equation (55) becomes

$$\log (T^L(t) - \log T^U(t)) = -2i \tan^{-1} (\pi t U(t) / T_c(t)) \qquad (56)$$

Hence equation (51) with equation (56) becomes

$$\log Q_+(z) = \int_0^1 \theta(t) \, dt / (t-z) \qquad (57)$$

where $\theta(t) = \pi^{-1} \tan^{-1} (\pi t U(t) / T_c(t)) \qquad (58)$

$$Q_+(z) = \exp \left( \int_0^1 \theta(t) \, dt / (t-z) \right) \qquad (59)$$

Equation (57) therefore indicates that $\log Q_+(z)$ and hence $Q_+(z)$ is analytic in the z plane cut along $(0,1)$ including infinity and has no zero in the z plane cut along $(0,1)$

We shall now find a relation between $Q_+(z)$ and $Q_-(z)$. In equation (45) we replace z by -z and we get

$$\log Q_-(-z) = - \int_0^1 \theta(t)\, dt/(t-z) \qquad (60)$$

From equation (57) we get

$$\log Q_+(z) = \int_0^1 \theta(t)\, dt/(t-z) \qquad (61),$$

Equation (61) and (62) gives

$$\log Q_-(-z) = - \log Q_+(z) \qquad (62)$$

Equation (62) gives us

$$Q_+(z)\, Q_-(-z) = 1 \qquad (63)$$

In equation (59) replacing z by –z

we get, $\quad Q_+(-z) = \exp \int_0^1 \theta(t)\, dt/(t+z) \,) \qquad (64)$

From equation (64) and equation (60) we get

$$Q_+(-z)\, Q_-(z) = 1 \qquad (65)$$

Hence the solution of the equation (7) for H(z) exists and it is a new form of H(z) given by equations(46) read with equation (32) and (37).

We shall verify the relation $H(z) H(-z) = (T(z))^{-1}$. In equation (32) with equation (37) we replace z by -z to get

$$H(-z) = S^{-1/2} (1-z)\, Q_-(-z) / (1 - k z) \qquad (66)$$

Equation (32) with equation (66) and (63) gives

$$H(z)\, H(-z) = (1 - z^2)\, Q_-(z) / (S\, Q_+(z)(1 - k^2 z^2)) \qquad (67)$$

Using equation (8), (20) in (67) we get

$$H(z) \, H(-z) = = (1-z^2) / (1-k^2 z^2) \, Q(z) = (T(z))^{-1} \quad (68).$$

We shall get new representation of dispersion function, $T(z)$. Equation (32), with equation (66) gives

$$T(z) = S(1-k^2 z^2) \exp\left(2 \int_0^1 \theta(t) \, t \, dt / (t^2 - z^2)\right) / (1-z^2), \quad (69)$$

where z lie in the plane cut along (-1,1). Therefore, $T(z)$, in equations (67-68) is expressed as a product of two functions $1/Q_+(z)$ and $Q_-(z)$ of which $Q_+(z)$ is analytic in the z plane cut along (0,1) and $Q_-(z)$ is analytic in the z plane cut along (-1,0). Those functions do not have any zero in their domain of analyticity.
We shall now go for the uniqueness of the solution $H(z)$. We have shown that the solution $H(z)$ of the linear non homogeneous integral equation (7) exist and they satisfy the relation $H(z) \, H(-z) = (T(z))^{-1}$
in the z plane cut along (-1,1). We have to show that this solution is unique. That means that we have to show that this new $H(z)$ should satisfy the nonlinear non homogeneous integral equation of $H(z)$ in the complex z plane cut along (-1,0).
Equation (31) using equation (37) gives

$$S^{-1/2} (1-z) \left(1 + z \int_0^1 U(u) H(u) \, du / (u-z)\right) / ((1-kz) Q_+(z)) = 1 \quad (70)$$

After some rearrangement we get from equation (70)

$$z \int_0^1 U(u) H(u) \, du / (u-z) = -1 + (1-kz) Q_+(z) S^{1/2} / (1-z) \quad (71)$$

Replacing z by –z in equation (71) we get

$$z \int_0^1 U(u) H(u) \, du / (u+z) = 1 - (1+kz) Q_+(-z) S^{1/2} / (1+z) \quad (72)$$

Using equations (32), (37) & (66) in equation (72) we get

$$z \int_0^1 U(u) H(u) \, du / (u+z) = 1 - 1/H(z) \quad (73)$$

Equation (73) is the nonlinear non homogeneous integral equation (1) for the $H(z)$. Hence it is concluded that this new $H(z)$ in the complex z plane cut along (-1,0) is unique.

Following Chandrasekhar[1], Busbridge[9], and Abhyankar and Fymat [22] we can also say that in non conservative cases, there is another solution $H_1(z)$ of equation (7) as
$$H_1(z) = (1+kz) H(z) / (1-kz). \quad (73a)$$
but in conservative cases it has only one solution of $H(z)$.

Hence from equations (32), (37), (46), (47) and (15) we get
the new representation of the H- function in non conservative case $U_0 < 1/2$

$$H(z) = A^{-1/2} (1+z) \exp\left(-\int_0^1 \theta(t)\, dt / (t+z)\right) / (1+kz), \qquad (74)$$

where $\quad \theta(t) = \pi^{-1} \tan^{-1}(\pi t\, U(t) / T_c(t)) \qquad (75)$

Similarly from equation (32), (37), (46), (49) and (15) we get the new representation of the H- function in conservative case $U_0 = 1/2$

$$H(z) = C^{-1/2} (1+z) \exp\left(-\int_0^1 \theta(t)\, dt / (t+z)\right), \qquad (76)$$

where $\quad \theta(t) = (\pi)^{-1} \tan^{-1}(\pi t\, U(t) / T_c(t)) \qquad (77)$

We shall now derive some useful formulae for numerical evaluation of H(z) from this new expression. Both for conservative cases and non conservative cases we have, for large z,

$$\int_0^1 \theta(t)\, dt / (t+z) = \theta_0 / z - \theta_1 / z^2 + \theta_2 / z^3 - \theta_3 / z^4 - - \qquad (78)$$

where $\quad \theta_r = \int_0^1 \theta(x)\, x^r\, dx, \quad r = 0,1,2,3 \text{-----} \qquad (79)$

$\theta_r$ are called the moments of $\theta(t)$ functions..

$$\exp\left(-\int_0^1 \theta(t)\, dt / (t+z)\right) = \exp(-\theta_0/z + \theta_1/z^2 - \theta_2/z^3 + \theta_3/z^4 - -\infty)$$

$$= 1 - s_{-1}/z + s_{-2}/z^2 - s_{-3}/z^3 + s_{-4}/z^4 - - \infty \qquad (80)$$

where $s_{-1} = \theta_0$;
$s_{-2} = \theta_0^2 / 2 + \theta_1$;
$s_{-3} = \theta_2 + \theta_0 \theta_1 + \theta_0^3 / 6$;
$s_{-4} = \theta_3 + \theta_0 \theta_2 + \theta_1^2/2 + \theta_0^2 \theta_1 /2$; etc $\qquad (81)$

In equation (81) all $\theta_r$ are required to be calculated

<u>In non conservative cases $U_0 < \tfrac{1}{2}$</u>, for large z,

$$(1+z)/(1+kz) = k^{-1} - \epsilon_2/z + \epsilon_3/z^2 - \epsilon_4/z^4 - -\infty \qquad (82)$$

where $\quad \epsilon_r = (k^{-r} - k^{-(r-1)}), \; r = 2,3,4, \text{------} \qquad (83)$

$$H(z) = H_0 - H_{-1}/z + H_{-2}/z^2 - H_{-3}/z^3 + H_{-4}/z^4 \ldots \infty \quad (84)$$

where $H_0 = A^{-1/2} k^{-1}$ ;

$$H_{-1} = A^{-1/2} (s_{-1} k^{-1} + \epsilon_2) ;$$

$$H_{-2} = A^{-1/2} (s_{-2} k^{-1} + s_{-1} \epsilon_2 + \epsilon_3) ;$$

$$H_{-3} = A^{-1/2} (s_{-3} k^{-1} + s_{-2} \epsilon_2 + s_{-1} \epsilon_3 + \epsilon_4) ;$$

$$H_{-4} = A^{-1/2} (s_{-4} k^{-1} + s_{-3} \epsilon_2 + s_{-2} \epsilon_3 + s_{-1} \epsilon_4 + \epsilon_5) ; \text{ etc} \quad (85)$$

For large $z$ we can determine $1/H(z)$, in non conservative cases $U_0 < \frac{1}{2}$, as

$$1/H(z) = p_0 + p_{-1}/z + p_{-2}/z^2 + p_{-3}/z^3 + \ldots \infty \quad (86)$$

wherefrom using $H(z) (H(z))^{-1} = 1$ we get

$$\begin{aligned} p_0 H_0 &= 1 ; \\ p_0 H_{-1} &= p_{-1} H_0 ; \\ p_0 H_{-2} &= p_{-1} H_{-1} - p_{-2} H_{-1} ; \\ p_0 H_{-3} &= p_{-1} H_{-2} - p_{-2} H_{-1} + p_{-3} H_0 ; \text{ etc} \end{aligned} \quad (87)$$

If $H_0, H_{-1}, H_{-2}, H_{-3} \ldots$ are determined from equation (85) with equation (81) and (82), then $p_0, p_{-1}, p_{-2}, p_{-3}, \ldots$ can be determined from equation (87).

We also know that $(H(-z))H(z))^{-1} = T(z)$ \quad (88)

For large $z$, we can expand both sides of equation (88) using equation (86),

$$(H(-z))H(z))^{-1} = t_0 - t_{-2}/z^2 - t_{-4}/z^4 - - - \infty , \quad (89)$$

where

$$\begin{aligned} t_0 &= p_0^2 ; \\ t_{-2} &= -2 p_0 p_{-2} + p_{-1}^2 ; \\ t_{-4} &= -p_0 p_{-4} + 2 p_{-3} p_{-1} - p_{-2}^2 ; \text{ etc} \end{aligned} \quad (90)$$

For large $z$,

$$T(z) = D - 2 U_2/z^2 - 2 U_4/z^4 - \ldots \infty \quad (91)$$

From equation (89) and (91) we get

$$\begin{aligned} D &= t_0 ; \\ 2 U_2 &= t_{-2} ; \\ 2 U_4 &= t_{-4} ; \text{ etc} \end{aligned} \quad (92)$$

. Using equation (74) we get, in non conservative cases,

$$H(z) H(-z) = (1 - z^2) A \exp(-2 \int_0^1 t \theta(t) dt / (t^2 - z^2)) / (1 - k^2 z^2) \quad (93)$$

For large $z$, in non conservative cases,

$$-2 \int_0^1 t \theta(t) dt / (t^2 - z^2) = 2\theta_1 / z^2 + 2\theta_3 / z^4 + 2\theta_5 / z^6 + ..\infty \quad (94)$$

$$H(z) H(-z) = \delta_0 + \delta_2 / z^2 + \delta_4 / z^4 + ......\infty \quad (95)$$

where

$$\delta_0 = H_0^2 ;$$

$$\delta_2 = 2 H_0 H_{-2} - H_{-1}^2 ;$$

$$\delta_4 = H_0 H_{-4} - 2 H_{-1} H_{-3} + H_{-2}^2 ; \text{ etc} \quad (96)$$

$$(1 - k^2 z^2) / (1 - z^2) = k^2 + (k^2 - 1) / z^2 + (k^2 - 1) / z^4 + .\infty \quad (97)$$

$$\exp(-2 \int_0^1 t \theta(t) dt / (t^2 - z^2)) = 1 + 2\theta_1 / z^2 + 2(\theta_3 + \theta_1^2) / z^4 + .\infty \quad (98)$$

$$H(z) H(-z) (1 - k^2 z^2) / (1 - z^2) = A^{-1} \exp(-2 \int_0^1 t \theta(t) dt / (t^2 - z^2)) \quad (99).$$

On comparing the like powers of $z^{-r}$ from both sides of equation (99)

$$k^2 \delta_0 = A^{-1} ;$$

$$k^2 \delta_2 + (k^2 - 1) \delta_0 = 2 \theta_1 A^{-1} ;$$

$$k^2 \delta_4 + (k^2 - 1) \delta_2 + (k^2 - 1) \delta_0 = 2 (\theta_3 + \theta_1^2) A^{-1} ; \text{ etc} \quad (100)$$

Using equation (84) and equation (91) we can get from

$$H(z) H(-z) T(z) = 1 \quad (101)$$

For large $z$, equating like powers of $z^{-r}$ from both sides of equation (101),

$$D \delta_0 = 1 ;$$
$$D \delta_2 = 2 U_2 \delta_0 ;$$

$$D \; \delta_4 = 2\delta_2 U_2 + 2 \delta_0 \; U_4; \text{ etc} \qquad (102)$$

Hence from equation (100) and equation (102), after some manipulation, we derive,

$$D = A \; k^2 \;;$$

$$\theta_1 = U_2 / D + ( k^2 - 1 ) / 2 k^2$$

$$\theta_3 = U_2^2 / D^2 + U_4 / D + ( k^4 - 1 ) / 4 k^4, \text{ etc} \qquad (103)$$

If we put $z = 0$ in equation (99) we get,

$$\theta_{-1} = -\tfrac{1}{2} \ln A \qquad (104)$$

For large $z$,

$$M(z) = z \int_0^1 U(x) H(x) \, dx / (x + z) = \alpha_0 - \alpha_1 / z + \alpha_2 / z^2 - \alpha_3 / z^3 -$$

$$1 / H(z) = 1 - M(z) = 1 - \alpha_0 + \alpha_1 / z - \alpha_2 / z^2 + \alpha_3 / z^3 - -- \qquad (105)$$

where $$\alpha_r = \int_0^1 x^r U(x) H(x) \, dx, \; r = 0,1,2,3 -- \qquad (106)$$

$$\alpha_0 = \int_0^1 U(x) H(x) \, dx = 1 - ( 1 - 2 U_0 )^{1/2} \qquad (107)$$

Here $\alpha_r$ are called the moments of H- functions.

From equation (105) and equation (86) we get, moments in non conservative cases,

$$\alpha_0 = 1 - p_0 \;;\; \alpha_1 = p_{-1} \;;\; \alpha_2 = - p_{-2} \;;\; \alpha_3 = p_{-3} \quad \text{etc.} \qquad (108)$$

In non-conservative case, we conclude that we have to calculate moments of $\theta$-functions, $\theta_{-1}, \theta_0, \theta_1, \theta_2, \theta_3$ .......from the expression (75) of $\theta(t)$ by way of integration   Out of those $\theta_{-1}, \theta_1, \theta_3$ .., odd moments are determinable from the exact relations (103) and (104) when zeros of $T(z)$ and moments of $U(x)$ are known, after calculation of $\theta_0, \theta_2, \theta_4$, we can calculate $s_{-1}$, $s_{-2}, s_{-3}, s_{-4}$, from equation (81); after calculation of $s_{-1}, s_{-2}, s_{-3}, s_{-4}, \ldots\ldots$, we can determine $H_0, H_{-1}, H_{-2}, H_{-3}$ ....... from equation (85); after determination of $H_0$, $H_{-1}, H_{-2}, H_{-3}$ ....... we can determine $p_0, p_{-1}, p_{-2}, p_{-3}, \ldots\ldots$ from equation (87); after determination of determine $p_0, p_{-1}, p_{-2}, p_{-3}, \ldots$ we can determine $t_0, t_{-2}, t_{-4}$, ....... from equation (90); after determination of $t_0, t_{-2}, t_{-4}$,   we can check those $t_0, t_{-2}, t_{-4}, \ldots\ldots$ with the exact values given by equation (92); after determination of $H_0, H_{-1}, H_{-2}, H_{-3}$ ......from Equation (85). we can determine $\delta_0, \delta_2, \delta_4, \ldots\ldots$ from equation (96); after determination of $\delta_0, \delta_2, \delta_4$. we can test the values $\theta_0, \theta_1, \theta_{-3}$, with the results obtained from equations (103) & (104). After determination of $p_0, p_{-1}$

, $p_{-2}$, $p_{-3}$, ……from equation (87). we can determine $\alpha_0$, $\alpha_1$, $\alpha_2$, $\alpha_3$ ……. from equation (105); Hence moments of H – function can be determined.

In conservative case, we follow the same procedure of non conservative cases as follows:

For large z, in conservative case, we get from equation (76),(81)

$H(z) = h_1 z + h_0 + h_{-1}/z + h_{-2}/z^2 + h_{-3}/z^3 + h_{-4}/z^4 - \infty$

$H(-z) = -h_1 z + h_0 - h_{-1}/z + h_{-2}/z^2 - h_{-3}/z^3 + h_{-4}/z^4 - \infty$

$1/(1-z^2) = -1/z^2 - 1/z^4 - 1/z^6 - \cdots \infty$

$H(z) H(-z) = \beta_0 - h_1^2 z^2 + \beta_2/z^2 + \beta_4/z^4 \ldots \infty$ (109)

where

$$\beta_0 = h_0^2 - 2 h_1 h_{-1},$$

$$\beta_2 = 2 h_0 h_{-2} - h_{-1}^2 - 2 h_1 h_{-3};$$

$$\beta_4 = h_{-2}^2 + 2 h_0 h_{-4} - 2 h_{-1} h_{-3} - 2 h_1 h_{-5}; \text{ etc} \quad (110)$$

Equation(109) with equation (81) gives

$h_1 = 1$,
$h_0 = 1 - s_{-1}$
$h_{-1} = s_{-2} - s_{-1}$;
$h_{-2} = s_{-3} - s_{-2}$;
$h_{-3} = s_{-4} - s_{-3}$;
$h_{-4} = s_{-5} - s_{-4}$; (110a)

where $s_{-1}$, $s_{-2}$, $s_{-3}$ …. are given by equation (81) and

$H(z) H(-z) / (1 - z^2) = h_1^2 + r_2/z^2 + r_4/z^4 + \ldots \infty$ (111)

where

$$r_2 = h_1^2 - \beta_0;$$

$$r_4 = h_1^2 - \beta_2 - \beta_0; \text{ etc} \quad (112)$$

For large values of z, in conservative case, we have,

$$H(z) H(-z) / (1 - z^2) = C^{-1} \exp\left(-2 \int_0^1 t \theta(t) \, dt / (t^2 - z^2)\right) \quad (113)$$

We compare the like powers of $z^{-r}$ from both sides of equation (113)

after using equation (98) and (111)) we get

$$C h_1^2 = 1;$$

$$r_2 C = 2 \theta_1$$

$$r_4 C = 2(\theta_3 + \theta_1^2); \text{ etc} \tag{114}$$

For large z, in conservative cases, we have

$$T(z) = -2 U_2/z^2 - 2 U_4/z^4 - 2U_6/z^6 \ldots \infty \tag{115}$$

$$H(z) H(-z) T(z) = 1 \tag{116}$$

and on equating power of $z^{-r}$ from both sides,

$$2 h_1^2 U_2 = 1;$$

$$h_1^2 U_4 = U_2 \beta_0;$$

$$h_1^2 U_6 = U_2 \beta_2 + U_4 \beta_0; \text{ etc} \tag{117}$$

Using equation (114) and (117) we can determine

$$\theta_{-1} = -1/2 \ln(C), \quad C = 2 U_2;$$

$$\theta_1 = 1/2 - U_4/2 U_2$$

$$\theta_3 = 1/4 + U_4/4 U_2^2 - U_6/2U_2, \tag{118}$$

We can also determine $\theta_1$ and $\theta_3$ in conservative cases from non conservative case, when k → 0, as a limiting value, When k→0 in T(1/k) =0 we get

$$\lim_{k \to 0} (D/k^2) = 2 U_2;$$

$$\lim_{k \to 0} (D - 2 k^2 U_2)/k^4 = 2 U_4; \tag{119}$$

Equation (103) gives, when k→0, using equation(119)

$$\theta_1 = 1/2 - U_4/2U_2; \tag{120}$$

We know that, when k is small, from T(1/k) = 0;

$$D = 2U_2 k^2 + 2U_4 k^4 + 2U_6 k^6 + \ldots\ldots \infty \qquad (121)$$

$$2U_2 / D = 1/k^2 - U_4/U_2 + (U_4^2/U_2^2 - U_6/U_2)k^2 + O(k^4) \quad , \quad (122)$$

Equation (121) and (122) gives

$$\lim_{k \to 0} (1/D - 1/2U_2 k^2) = -U_4/2U_2^2 \qquad (123)$$

$$\lim_{k \to 0} 2U_2 (1/Dk^2 - 1/2U_2 k^4 + U_4/2U_2^2 k^2) = (U_4^2/U_2^2 - U_6/U_2) ; \quad (124)$$

$\theta_3$ can be written as

$$\theta_3 = (U_2/D - 1/2k^2)(U_2/D + 1/2k^2) + U_4/D + 1/4 , \qquad (125)$$

Using equation (122), we get

$$\lim_{k \to 0} (U_2/D - 1/2k^2)(U_2/D + 1/2k^2) = \\ -U_4/2U_2 k^2 + (U_4^2/U_2^2 - U_6/U_2)/2 + \\ + U_4^2/4U_2^2 + O(k^2) ; \qquad (126)$$

$$\lim_{k \to 0} U_4/D = U_4/2U_2 k^2 - U_4^2/2U_2^2 + O(k^2) ; \qquad (127)$$

Using equation (126), (127) in equation (125) we get

$$\theta_3 = 1/4 + U_4/4U_2^2 - U_6/2U_2 ; \qquad (128)$$

We, in conservative case, therefore, also conclude that

From equation (75) we have to determine $\theta_{-1}, \theta_0, \theta_1, \theta_2, \theta_3$ …….from the expression of $\theta(t)$ by way of integration. Out of those $\theta_{-1}, \theta_1, \theta_3$ .., odd moments are determinable from the exact relations equation (118) when the moments of U(x) are known. After calculation of $\theta_0, \theta_2, \theta_4,\ldots\ldots$ we can calculate $s_{-1}, s_{-2}, s_{-3} \ldots$ from equation (81) with $U_0 = \frac{1}{2}$, after calculations of $s_{-1}, s_{-2}, s_{-3} \ldots$ we can determine $h_0, h_1, h_{-1}, h_{-2}, h_{-3} \ldots\ldots$ from equation (110a) ; after determination of $h_0, h_1, h_{-1}, h_{-2}, h_{-3}$ …we shall determine $\beta_0, \beta_2$ etc from equation (110). Thereafter determining $r_2, r_4,$ etc……… from equation (112) .we can check the result with the exact value obtained from relation (114).

We know that, in conservative case,

$$1/H(z) = 1 - M(z) = \alpha_1/z - \alpha_2/z^2 + \alpha_3/z^3 - \text{--} \quad \text{when z is large} \quad (129)$$

22where $\quad \alpha_r = \int_0^1 x^r U(x) H(x) \, dx \quad , \quad r = 1,2,3 \; - - \quad (130)$

$$\alpha_0 = \int_0^1 U(x) H(x) \, dx = 1 \quad (131)$$

Here $\alpha_r$ are called the r th moment of H- functions.

In conservative case, for large z, we shall get

$1/H(z) = n_{-1}/z + n_{-2}/z^2 + n_{-3}/z^3 + \text{-----} \infty \quad (132)$

but $\quad H(z) \, (H(z))^{-1} = 1$ gives

$\quad n_{-1} = 1/h_0 ;$

$\quad n_{-2} = -h_1/h_0 ;$

$\quad n_{-3} = h_1^2/h_0^2 - h_{-1}/h_0^2 ; \text{ etc} \quad (133)$

and

$\quad \alpha_1 = n_{-1} \; ; \; \alpha_2 = -n_{-2} \; ; \; \alpha_3 = n_{-3} \quad (134)$

Hence after determination of $h_0$, $h_1$, $h_{-1}$, $h_{-2}$, $h_{-3}$ ……. we can determine $n_{-1}$, $n_{-2}$, $n_{-3}$, ……. from equation (133) ; after determination of $n_{-1}$, $n_{-2}$, $n_{-3}$, ……. we can determine $\alpha_1$, $\alpha_2$, $\alpha_3$ ……. from equation (134). Hence moments of H – function in conservative case can be determined
.
Conclusion : This Wiener-Hopf procedure for deriving the H-functions from linear non homogeneous integral equation may be used in solving equations of transfer for problems of anisotropic and non coherent scattering both for finite and infinite atmosphere with Laplace transform . The numerical evaluation of the H-function for a simple U(u) is made ready and awaiting communication .

Acknowledgement .I express my sincere thanks to the Department of Mathematics , Heritage Institute of Technology , Anadapur, West Bengal , India for their whole hearted support .

**The End.**

.